\documentclass[sigconf]{acmart}

\usepackage{tabularx}

\usepackage{tikz}
\usepackage{graphicx}
\usepackage{subcaption}
\usepackage{makecell}

\begin{document}

\title{Learning Cross-Domain Media Similarity from Amazon Reviews}

\author{Gabrielly Xavier dos Santos, Lucas Dias Cler, Luciana Fujii Pontello, Olga Goussevskaia}
\email{gabrielly.santos@dcc.ufmg.br, lucasdiascler@dcc.ufmg.br, lucianafujii@gmail.com, olga@dcc.ufmg.br}


\affiliation{
  \institution{Computer Science Department, Universidade Federal de Minas Gerais (UFMG)}
  \city{Belo Horizonte}
  \country{Brazil}
  \postcode{31270-901}
}

\begin{abstract}
Collaborative filtering is a basic foundation of many recommendation algorithms. It works by analyzing a user's history and then suggesting items similar to those in that history. This method typically relies on user data to compute item similarity, data that is widely available to commercial systems but is usually sparse or inaccessible for independent systems and research, especially across multiple media domains.

In this work, we propose a method to generalize similarity information between items using supervised learning. As input to our model, we use semantic and temporal features extracted from Amazon product reviews, which are publicly available. As the training target, we use item (product) co-occurrence in the reviewing history of Amazon customers. We evaluate the quality of the predicted similarities across two media domains: books and movies. Experimental results demonstrate the robustness of our methodology, yielding positive outcomes in zero-shot cross-domain transfer tasks. Furthermore, feature attribution analysis reveals a structural shift in how feature groups are utilized across setups: intra-domain models rely heavily on linguistic semantics, whereas cross-domain environments pivot toward behavioral popularity cues. Incorporating even limited target-domain context during training bridges this domain sparsity gap, significantly enhancing framework generalization.
\end{abstract}

\keywords{Multimedia, similarity, data sparsity, multi-domain, cross-domain, recommendation algorithms}

\maketitle

\section{Introduction}
\label{sec:intro}

Personalized recommendation has become one of the primary mechanisms through which users discover content across e-commerce, media streaming, and publishing platforms. The dominant paradigm underlying these systems is collaborative filtering (CF), which leverages collective user behavior to suggest items similar to a user's interaction history. Specifically, item-to-item CF computes similarity directly from co-occurrence patterns in user histories (e.g., items frequently co-purchased or co-consumed), serving as a foundational production technique since its inception~\citep{linden2003item}. Latent-factor approaches, such as matrix factorization, later generalized this principle by optimizing low-dimensional representations to reconstruct these observed interactions~\citep{koren2009matrix}.

However, a structural limitation of these methods is their absolute dependence on massive user interaction data. While abundant for major commercial platforms, such granular logs are rarely accessible to independent developers or academic researchers. This data asymmetry creates a significant barrier between industrial practice and academic research, confining the latter to static datasets that fail to reflect real-world production environments~\citep{zou2025survey}, while severely hindering experimental verification and reproducibility~\citep{dacrema2021troubling}. This gap is further compounded in cross-domain settings, where similarities must be mapped across distinct media types (e.g., books and movies) and where the cross-domain user overlap required by traditional methods is frequently non-existent~\citep{zhang2025crossdomain}.

\textbf{State of the art in similarity learning.}
Recent literature addresses this sparsity bottleneck through two primary paradigms. The first augments interaction signals with item content, demonstrating that semantic features extracted from user-generated review texts can partially substitute for missing interaction histories~\citep{zheng2017joint}. Large review corpora, such as the Amazon dataset~\citep{ni2019justifying}, have democratized this content-based approach, even though the underlying co-purchase graphs remain strictly proprietary. 

The second paradigm transfers similarity structures across domains, either by aligning user representations across overlapping categories~\citep{zhao2023crossdomain,cao2023universal,zhao2024discerning} or by utilizing public editorial metadata as a proxy for private usage data~\citep{alonsojimenez2022music}. Alternatively, some approaches bypass representation learning entirely, framing relationship estimation as a link-prediction task over product graphs~\citep{liu2025amazon}. Across these directions, the core strategy relies on using public proxies—text, metadata, or graphs—to approximate a similarity signal otherwise locked behind private commercial infrastructure.

\textbf{Novelty and contributions.}
Distinct from prior literature, we do not treat the proprietary co-occurrence signal as fundamentally unavailable, but rather as a concrete, well-defined supervised learning target. Instead of relying on implicit correlations via metadata-driven representations~\citep{alonsojimenez2022music} or assuming partial user overlaps~\citep{zhang2025crossdomain}, we directly regress onto item co-occurrences derived from proprietary histories, using exclusively semantic and temporal features extracted from public reviews. 

This reframes the data-sparsity problem as a supervised generalization task: a model trained where both signals are observed can estimate high-quality similarity wherever public features exist, bypassing the need for private logs. We validate this framework across two distinct media domains (books and movies), demonstrating cross-modality generality. Our contributions are fourfold:

\begin{enumerate}
    \item We formulate an item co-occurrence prediction as a supervised learning task over public review text and temporal signals, enabling production-grade similarity estimation without private interaction data.
    \item We design and evaluate a framework that generalizes this similarity signal across heterogeneous media domains, proving its resilience across distinct content modalities.
    \item We present a feature attribution analysis via SHAP \cite{lundberg2017shap} and block-level feature ablation, uncovering a structural shift where intra-domain tasks rely heavily on text semantics, while cross-domain generalization pivots toward behavioral popularity cues.
    \item We provide a comprehensive empirical evaluation against true proprietary co-occurrence data, offering independent researchers a concrete, reproducible alternative to inaccessible commercial datasets.
\end{enumerate}

We validate our approach on the \texttt{Amazon Reviews 2023} dataset \citep{hou2024bridging}, from which we extracted two distinct media domains, \texttt{BOOKS} and \texttt{MOVIES}, comprising together over 5M items, 46M reviews, and 16M users. Our methodology establishes the target signal (ground truth) as the structural co-occurrence frequency within item pairs (i.e., the volume of shared users who actively reviewed both items). 
Beyond the global statistics, we focused on analyzing the cross-domain item pairs, as these connections provide the essential signal to bridge the semantic gap between categories. 
We identified  $\approx$1.3M cross-domain pairs, representing 35.0\% of the (post-filtered) dataset. 

Our experiments demonstrate that the primary challenge lies in executing a zero-shot domain transfer from a homogeneous intra-domain environment (book-book and movie-movie pairs), governed by highly localized feature rules, to a heterogeneous cross-domain environment (book-movie pairs) with distinct underlying dynamics. 
In fact, the \textit{intra-domain $\rightarrow$ cross-domain} transfer learning setup results in a negative $R^2$ score (-4.00211). This demonstrates that the decision rules learned exclusively from intra-domain pairs fail entirely to generalize to cross-domain interactions.
However, our results also reveal that introducing even a modest portion of cross-domain examples into the training phase significantly enhances model adaptability. Despite being disproportionately smaller than the volume of intra-domain data, including these cross-domain pairs effectively bridges the domain sparsity gap. 

Finally, we evaluated the performance of our framework under classical transfer learning regimes.
We partitioned the intra-domain pool into \texttt{BOOKS} and \texttt{MOVIES}. Our empirical results demonstrate strong generalization capabilities, yielding a positive $R^2$ of 0.38 for the \texttt{BOOKS} $\rightarrow$ \texttt{MOVIES} pipeline and an exceptional $R^2$ of 0.76 for the \texttt{MOVIES} $\rightarrow$ \texttt{BOOKS} transition. This high performance explicitly confirms that substantial underlying similarities exist across these distinct structural domains. 


The rest of the paper is organized as follows. Section~\ref{sec:related} reviews cross-domain and content-based recommendation literature. Section~\ref{sec:dataCollection} outlines our data collection pipeline, while Section~\ref{sec:dataCharacterization} details the data characterization and core-filtering protocols. Section~\ref{sec:method} details our supervised architecture and feature extraction layers. Section~\ref{sec:results} presents quantitative results, ablation studies, and interpretability insights. Finally, Section~\ref{sec:conclusion} concludes the paper and outlines future research directions.
\section{Related Work}
\label{sec:related}


Our work sits at the intersection of three lines of research: (i) cross-domain
recommendation, which seeks to transfer relevance signals across item domains;
(ii) content- and review-based recommendation, which substitutes textual side
information for interaction data; and (iii) the broader, recently-documented
gap between the data available to academic researchers and the data used in
industrial recommender systems. We discuss each in turn and position our
contribution relative to them.

\textbf{Cross-Domain Recommendation.}
A large body of work addresses the problem of transferring preference or
similarity signal from a data-rich source domain to a sparser target domain.
\citet{zhang2025crossdomain} provide a recent taxonomy of this area,
organizing methods according to how cross-domain connections are established
(overlapping users/items, shared features, or content) and how the resulting
information is fused for recommendation. Within this taxonomy, our setting is
closest to \emph{content-based} cross-domain transfer, since the media
domains we consider (books and movies) share no overlapping users or
items and must instead be bridged through item content.

Several recent systems instantiate this idea. \citet{zhao2023crossdomain}
align user interest representations across domains to mitigate the cold-start
problem for users with little interaction history in the target domain.
\citet{cao2023universal} propose a single, universal model capable of
transferring domain-shared information across a variety of cross-domain
scenarios without requiring domain-specific architectures.
\citet{zhao2024discerning} further refine cross-domain user representations
by explicitly disentangling domain-shared and domain-specific preference
signals. These approaches, however, are predominantly designed to transfer
\emph{user}-level preferences using collaborative or sequential interaction
data on at least one side of the transfer. In contrast, our method learns a
purely \emph{item}-level similarity function from public content features,
without assuming access to any interaction data, which makes it applicable
even when no user-level signal is available in either domain at training
time.

Furthermore, recent efforts in multi-domain contexts have focused on optimizing the learning process itself and aligning with multifaceted user profiles. For instance, \cite{23webmediaMultidomain} explore training optimization by proposing a batch-balancing strategy with domain information to mitigate the dominance of data-rich domains over sparser ones. From a personalization perspective, \cite{25webmediacalibration} address the diverse dimensions of user preferences through a multi-category calibration method, ensuring that recommendations align proportionally with the varied categories a user interacts with, moving beyond pure accuracy metrics.

\textbf{Review- and Content-Based Recommendation.}
Using review text or other item content as a substitute for sparse
interaction data has a long history in recommender systems research. A
foundational example is \citet{zheng2017joint}, who jointly model users and
items from review text using parallel convolutional networks, showing that
semantic features extracted from reviews can alleviate the cold-start and
data sparsity problems intrinsic to collaborative filtering. Our approach
follows this general strategy of treating review text as a public,
recoverable substitute for private interaction signals, but differs in its
supervised learning target: rather than predicting a rating or a
user-specific preference score, we train a model to regress onto item
co-occurrence in users' reviewing histories, a relational signal that is
itself privately held by the platform and not exposed by individual reviews.
This connects our work to recent efforts that recover relational structure
between items directly from public metadata, such as the co-purchase link
prediction system of \citet{liu2025amazon}, who construct a co-purchasing
graph from the Amazon Product Co-Purchasing Network Metadata and apply
graph neural methods to predict new co-purchase edges. Where that line of
work primarily leverages graph structure and item metadata, we instead rely
on semantic and temporal features extracted directly from review text,
without assuming access to an existing co-purchase graph at inference time.

\textbf{The Academia--Industry Data Gap.}
Our problem framing is motivated by a growing body of work that documents
the divide between the data available to academic researchers and the data
that underlies industrial recommender systems. \citet{zou2025survey}
characterize this divide directly, arguing that academic research remains
largely confined to offline, public datasets and lacks access to the
real-world interaction data that drives industrial systems, which limits the
practical relevance of academic progress. Relatedly,
\citet{dacrema2021troubling} show that a surprisingly large fraction of
recently published neural recommendation approaches fail to reproduce, or
are outperformed by simple baselines once properly tuned, a finding partly
attributable to inconsistent or inaccessible experimental data and code. Our
work can be read as a constructive response to this gap: rather than
treating proprietary co-occurrence data as a precondition for building
accurate item-similarity models, we show that it can be approximated, to a
useful degree, purely from publicly available review content.


\textbf{Similarity Learning from Public Metadata.}
The work in \cite{pontello2018direction} explored methodologies for computing item similarities across different media domains. A major difficulty highlighted in that work was the extreme data sparsity, which was mitigated by extracting semantic features from Wikipedia pages matching the items' (book, movies or music) titles. 
Our proposal differs from this methodology by leveraging Amazon user review data. This dataset not only allows us to incorporate explicit user ratings to formulate robust ground-truth metrics, but also significantly improves the feature representation by adding interaction-based temporal dynamics and aggregated statistics—such as review volume and item lifespan—extracted directly from the user reviews.

\section{Data Collection}
\label{sec:dataCollection}


\textbf{\texttt{Amazon Reviews 2023.}} For our experiments, we use the \texttt{Amazon Reviews 2023}
dataset~\citep{hou2024bridging}, a large-scale collection of customer reviews and item metadata covering 33 product categories on Amazon,
including the two media domains studied in this work: books and movies. The dataset was introduced as an updated successor to the
earlier Amazon Reviews releases~\citep{ni2019justifying}.
It was collected using a large-scale, user-centric crawling pipeline from publicly available Amazon pages. Reviews span from June 1996 to September 2023, covering over 50 million users and 50 million items (Table \ref{tab:domain_summary_combined}).

\textbf{Domain Extraction.} From the Amazon Reviews dataset we extracted four specific category subsets and established two distinct media domains: 
\begin{itemize}
    \item \texttt{BOOKS}: Using the ``Books`` reviews subset.
    \item \texttt{MOVIES}: Using the ``Movies and TV`` reviews subset.
\end{itemize}

\textbf{Feature Extraction.} Each raw domain dataset contains 10 original features: 'rating', 'title', 'text', 'images', 'asin', 'parent\_asin', 'user\_id', 'timestamp', 'helpful\_vote', and 'verified\_purchase'. To construct our predictive framework, we filtered and utilized a core subset of these metadata attributes: 'rating', 'title', 'text', 'parent\_asin', 'user\_id', and 'timestamp'. The remaining attributes (e.g., images, helpful\_vote, verified\_purchase) were discarded as they did not directly contribute to the core semantic or behavioral dimensions analyzed in this study.


\textbf{Preprocessing.}
To process tens of millions of raw interactions within local memory limits, the datasets were converted to columnar Parquet files, optimizing storage efficiency and preserving all metadata columns. We integrated DuckDB to perform high-throughput SQL operations directly on the Parquet files, executing selective column scans that drastically reduced I/O overhead and bypassed the in-memory limitations of standard row-based tools. Once the datasets were filtered to a manageable scale, they were transferred to Pandas for final Natural Language Processing (NLP) and machine learning engineering. Lastly, a de-duplication protocol was applied to eliminate redundant logs, retaining only the most recent interaction per user-item pair to capture final user preferences accurately.
\section{Data Characterization and Core-Filtering}
\label{sec:dataCharacterization}

\begin{figure}[t]
\centering
\includegraphics[width=1\columnwidth]{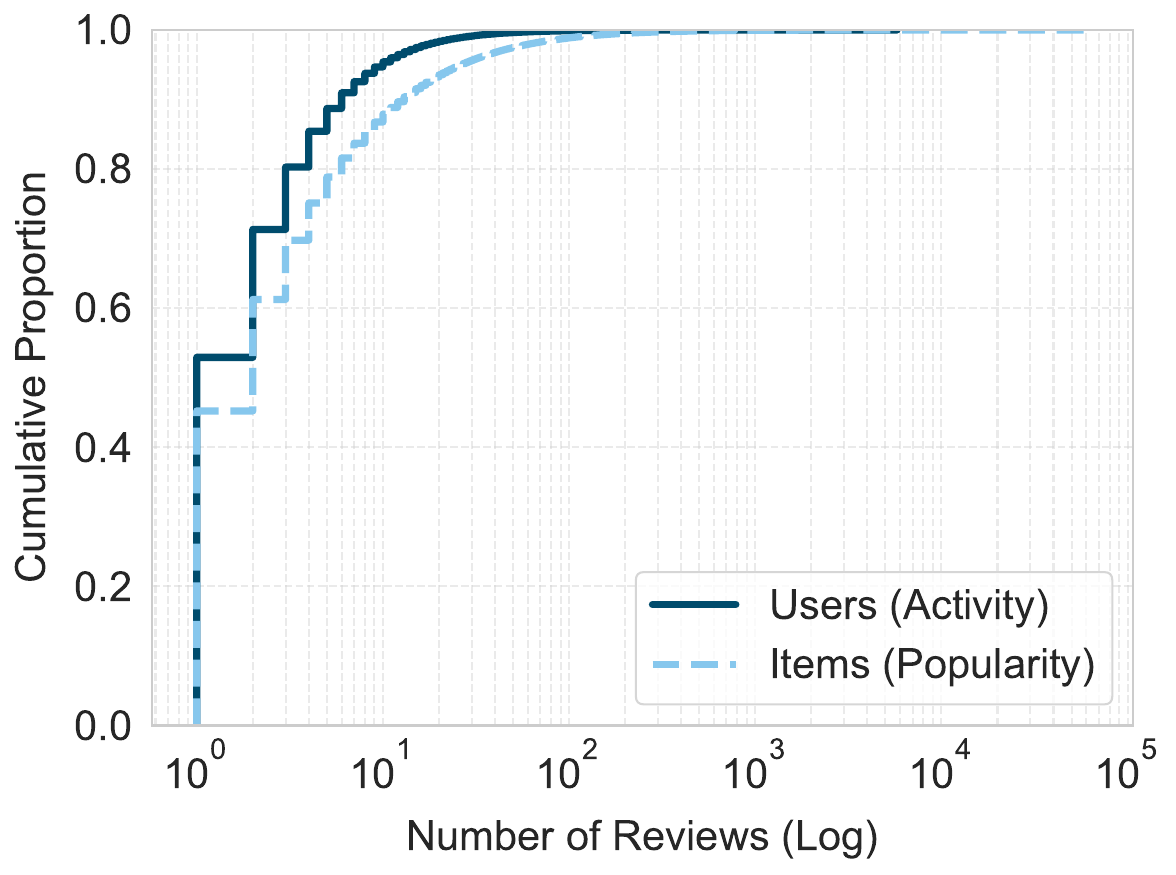}
\caption{ECDF of reviews per user and per item (before core-filtering).}
\label{fig:cdf_before}
\end{figure}

Figure~\ref{fig:cdf_before} presents the Global Empirical Cumulative Distribution Functions (ECDF) for both user interactions and item popularity. The initial distributions reveal a severe long-tail behavior and extreme sparsity: 80\% of the items received six reviews or fewer, and 80\% of the users made three reviews or fewer. Given our item-based collaborative filtering methodology \cite{linden2003amazon}, relying on such scarce interactions would introduce excessive statistical noise. 

To guarantee robust similarity representations, we applied a strict 10-core filtering threshold, retaining only items with 10 or more reviews. This filtering successfully shifted the global density, establishing a significantly healthier training distribution with a median of 20 reviews per item. Table~\ref{tab:domain_summary_combined} contrasts the dataset dimensions before and after this filtering phase.

\begin{table}[t] 
    \centering
    \caption{Dataset size before and after core-filtering.}
    \small
    \label{tab:domain_summary_combined}
    \vspace{0.1in} 
    \begin{tabular*}{\linewidth}{@{\extracolsep{\fill}} llrrr @{}}
        \hline
        \textbf{Domain} & \textbf{Filtering} & \textbf{Items} & \textbf{Reviews} & \textbf{Users} \\
        \hline
        \makecell{\texttt{Amazon}\\\texttt{Reviews 2023}}&  & 48,185,153 & 571,544,897 & 54,514,264 \\
        \hline
        \texttt{BOOKS}  & Before & 4,446,063 & 29,139,325 & 10,297,355 \\
                       &  $\ge 10$ reviews  &   502,048 & 20,024,781 &  8,183,738 \\
        \hline
        \texttt{MOVIES} & Before &   747,764 & 17,158,519 &  6,503,429 \\
                       & $\ge 10$ reviews  &   186,296 & 15,672,563 &  6,177,701 \\
        \hline
    \end{tabular*}
\end{table}

Furthermore, we mapped user intersection across categories to assess cross-domain viability. Within the raw total of 14,269,638 unique users, we isolated the \textit{Bridge Users}—individuals with active interactions spanning both domains—who serve as our anchor for cross-domain knowledge transfer. In the raw pool, Bridge Users accounted for 17.74\% of the population (2,531,146 users), confirming a dense overlapping signal. 

\begin{figure}[t] 
\centering
\includegraphics[width=1\columnwidth]{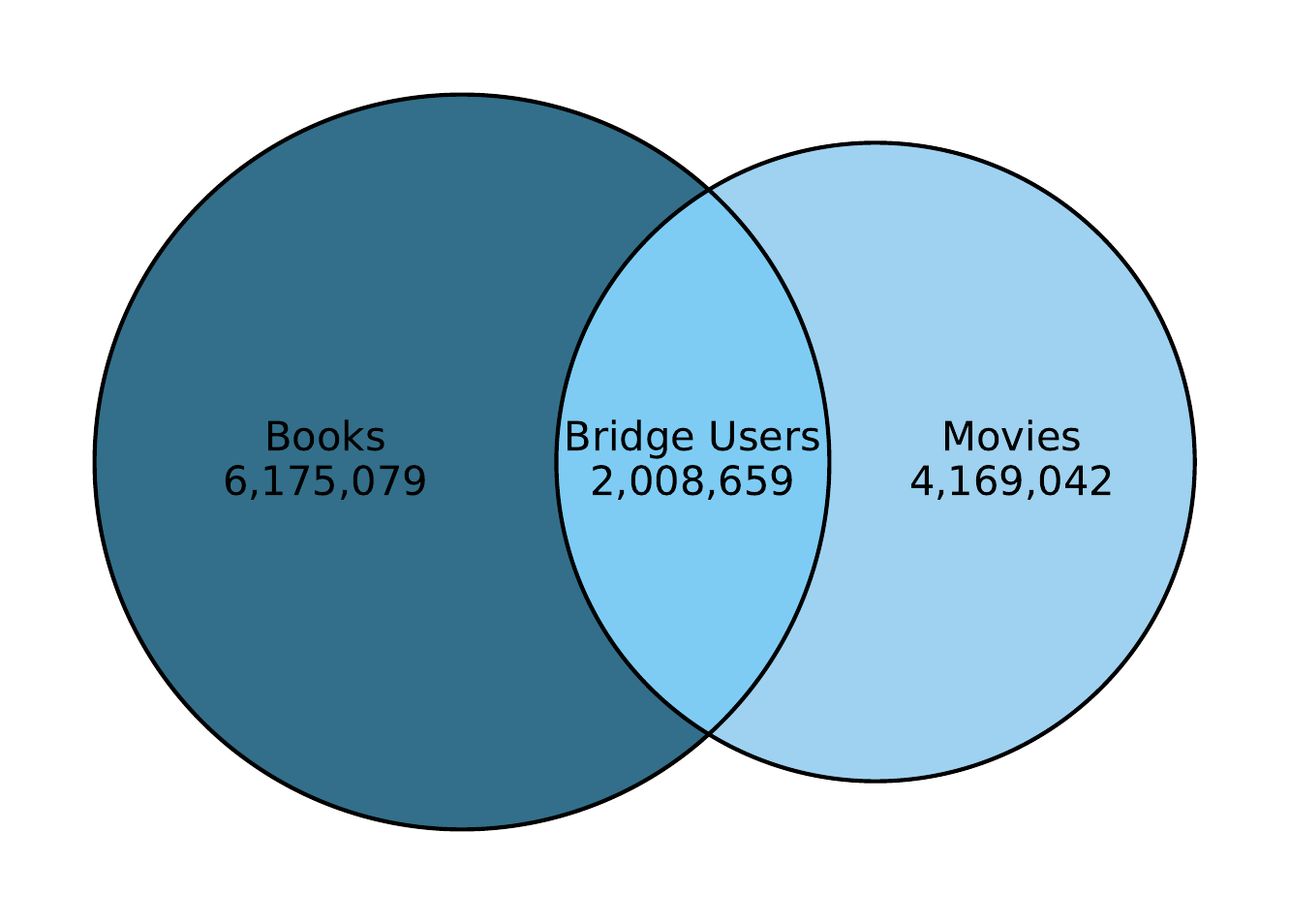}
\caption{\textit{Bridge users:} User intersection between \texttt{Books} and \texttt{Movies} domains after core-filtering.}
\label{fig:venn_clean}
\end{figure}

While the core-filtering step inevitably pruned the interaction graph—resulting in an approximate 20\% loss of Bridge Users (as visualized in the post-filter intersection in Figure~\ref{fig:venn_clean})—this trade-off was vital to safeguard the statistical integrity and robustness of the ground-truth formulation in the subsequent architectural phases.

\section{Similarity Learning: Methodology}
\label{sec:method}

The primary objective of our methodology is to develop a scalable supervised framework that predicts high-quality item similarity. By doing so, our system can generalize these similarity mappings across catalogs without depending on private or inaccessible industrial user logs. To achieve this, our approach is structured into three interconnected phases: first, we formulate a robust ground truth signal based on filtered item co-occurrence frequencies and calculate continuous item-to-item targets using cosine similarity; second, we design a multi-dimensional feature engineering pipeline to extract semantic embeddings from review texts alongside temporal, popularity, and quality metadata; and third, we integrate these distinct feature blocks into a unified hybrid vector to train and evaluate our predictive models across heterogeneous domain scenarios.

\subsection{Ground Truth Formulation}
\label{sec:groundTruth}

Our next primary goal is to formulate a reliable ground truth. In contrast to traditional recommendation frameworks that define ground truth based on isolated user-item interactions (such as explicit ratings or binary clicks), our methodology establishes the target signal as the structural co-occurrence frequency within item pairs (i.e., the volume of shared users who actively reviewed both items). This formulation is strictly necessary because our objective is to map cross-domain item-to-item similarity representations; therefore, evaluating the density of shared user behavior allows us to bypass the standard user-sparsity constraint and directly capture the underlying behavioral alignment between distinct media categories.

\begin{figure}[t] 
\centering
\includegraphics[width=1\columnwidth]{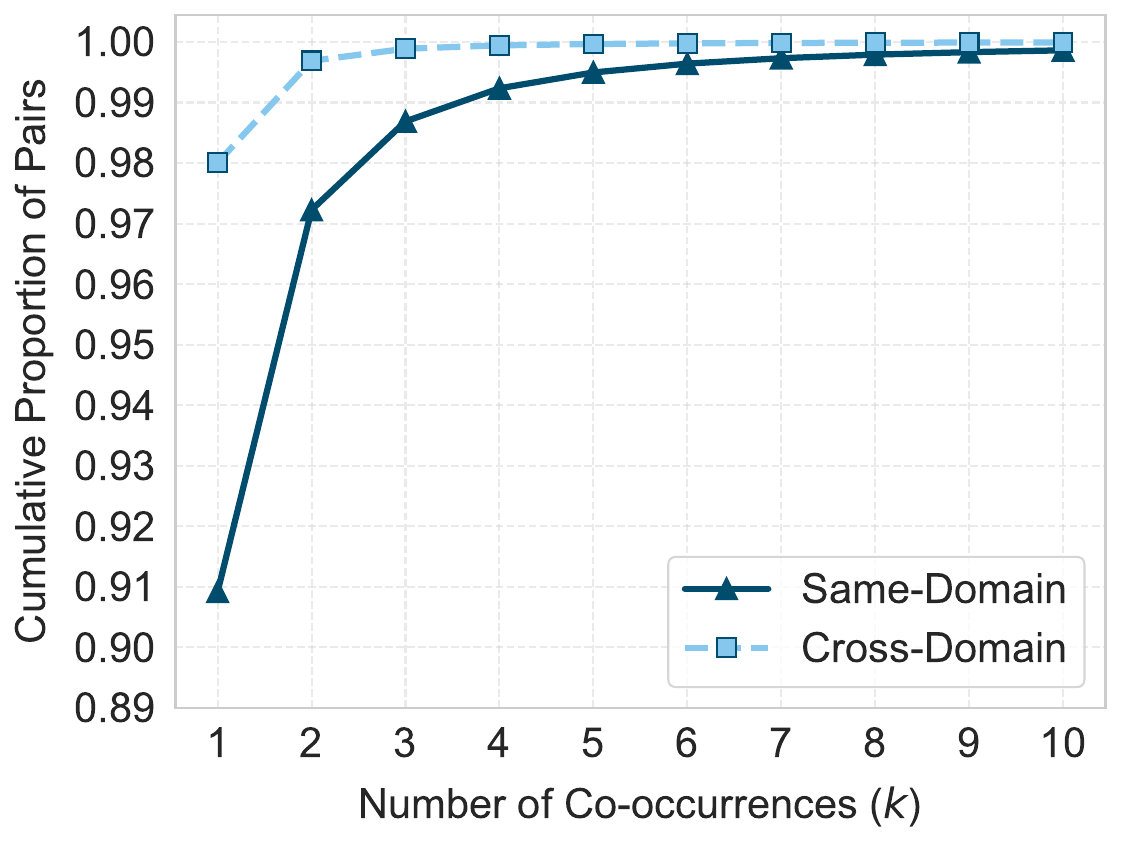} 
\caption{CDF of item pairs by co-occurrence frequency.}
\label{fig:cdf_pairs_coocor}
\end{figure}

As illustrated in Figure \ref{fig:cdf_pairs_coocor}, selecting a co-occurrence threshold ($k$) presents a clear trade-off between statistical significance and data volume. A higher $k$ ensures stable and reliable similarity scores backed by sufficient shared user behavior, whereas a lower $k$ yields more available pairs for training our models. 

We applied distinct co-occurrence thresholds ($k$) to ensure the reliability of our Ground Truth. For same-domain pairs, we used a filter of $k \ge 5$, effectively removing the bottom $99.5\%$ of sparse interactions to retain only highly confident similarity scores. However, given the natural sparsity of user intersections between different categories, applying this same filter to cross-domain pairs would result in an insufficient dataset for predictive modeling. Therefore, we opted for a threshold of $k \ge 2$ for cross-domain pairs, preserving enough data volume to train the machine learning algorithms while maintaining a baseline of shared user interactions.

\textbf{Cosine-based Similarity.}
To compute the similarity between item pairs, we adopted the item-based cosine similarity metric, a widely recognized approach in collaborative filtering \cite{SarwarKKR01}, applied similarly to the methodology described by \cite{pontello2018direction}. 

In this approach, each item is represented as a vector within the multidimensional space of users who have rated them. The similarity between any two items, $i$ and $j$, is quantified by calculating the cosine of the angle between their respective rating vectors, $\vec{i}$ and $\vec{j}$. Formally, the similarity score $sim(i,j)$ is defined as:
\begin{equation}
sim(i,j) = \cos(\vec{i}, \vec{j}) = \frac{\vec{i} \cdot \vec{j}}{\|\vec{i}\|_2 * \|\vec{j}\|_2},
\label{eq:cosine_formula}
\end{equation}

\begin{figure}[t] 
\centering
\includegraphics[width=1\columnwidth]{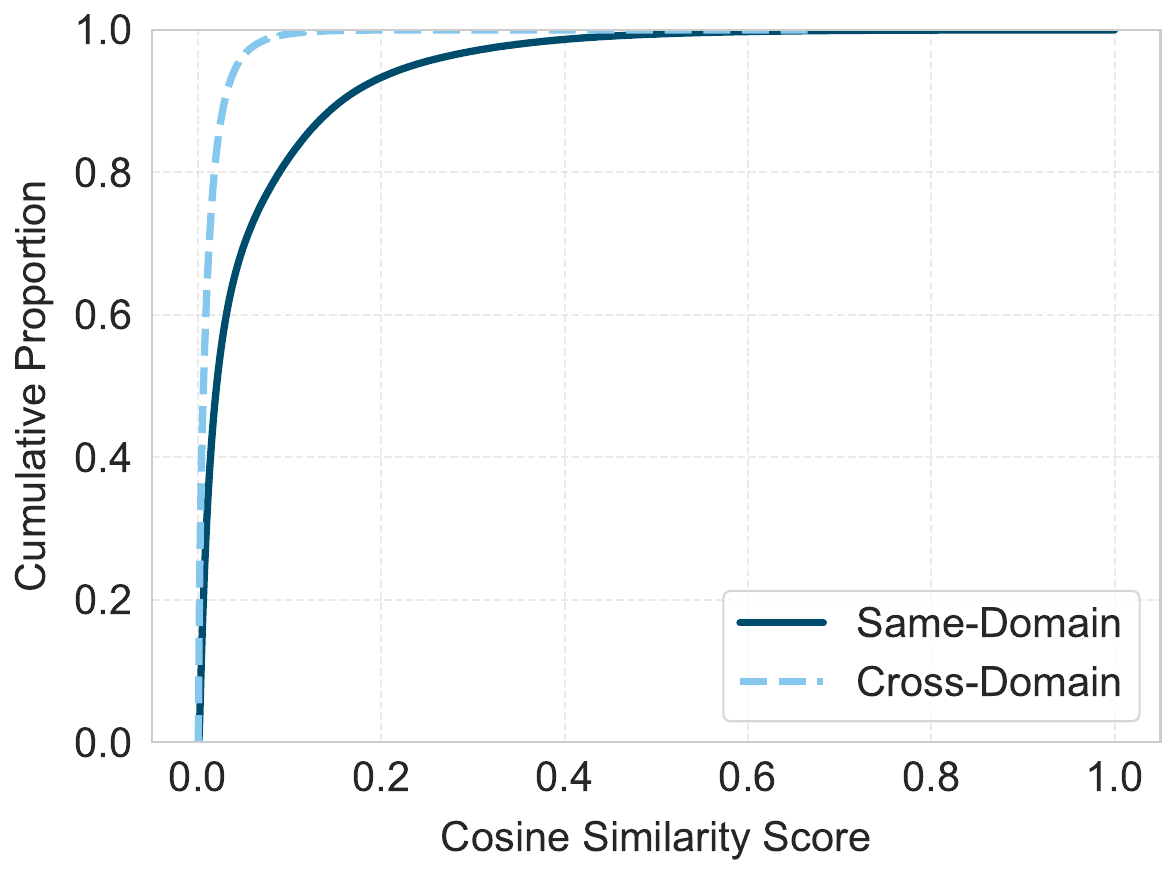} 
\caption{Cosine similarity distribution among same- and cross-domain item pairs.}
\label{fig:cdf_SameVsCross}
\end{figure}

Figure \ref{fig:cdf_SameVsCross} illustrates the distribution of the computed cosine similarity scores. As anticipated in highly sparse environments, a significant portion of the values is concentrated near zero.

Beyond the global statistics, our primary focus lies in analyzing the cross-domain pairs, as these connections provide the essential signal to bridge the semantic gap between categories. Out of the 3,878,976 total finalized pairs generated by our filtering pipeline, the intra-domain baseline is comprised of 1,450,872 Book-to-Book pairs (approximately 37\%) and 1,074,128 Movie-to-Movie pairs (approximately 28\%). Crucially, we identified exactly 1,353,976 heterogeneous cross-domain pairs, representing the remaining 35.0\% of the complete dataset. This substantial volume provides a rich distribution of cross-category interactions, which serve as the foundation for evaluating framework generalization.

\subsection{Feature Engineering}
\label{sec:features}

Data sparsity limits the accuracy of relying solely on user-item interactions to infer catalog similarities. To maximize information extraction from raw data, we design a pipeline divided into four primary dimensions: Semantic Encoding, Temporal Metrics, Popularity Metrics, and Ratings Metrics.

\subsubsection{Semantic Encoding}
Leveraging review text helps mitigate interaction sparsity by preserving rich context~\cite{v2025contrastive}. For the 231,346 unique items within our pairs, we implement a three-step aggregation and embedding workflow:

\begin{itemize}
    \item \textbf{Text Aggregation:} Review texts and titles are concatenated per item. To prevent Out-of-Memory (OOM) bottlenecks during large-scale string processing, we leverage DuckDB~\cite{raasveldt2019duckdb} for out-of-core batch computation in optimized chunks.
    \item \textbf{Keyword Extraction:} We apply TF-IDF~\cite{manning2008tfidf} to extract the top 50 most relevant keywords per product. Given our 10-core filtering threshold, this extraction serves as a robust statistical filter across the item catalog. Limiting the representation to the top 50 terms successfully isolates the core semantic essence of each item while systematically eliminating sparse or uninformative linguistic noise.
    \item \textbf{Embedding Generation:} These 50 keywords are fed into a pre-trained SBERT framework~\cite{reimers2019sentence} to generate dense item embeddings, lowering computational overhead while preserving high-quality semantic representations.
\end{itemize}

\subsubsection{Temporal Metrics}
To capture evolving user engagement while avoiding the computational bottlenecks of processing raw timestamps across highly skewed interaction volumes, we implement a temporal aggregation strategy based on the Jaccard Index \cite{2009korentemporal,pontello2018direction}. Each item is represented by its set of unique active years ($Y$). For a given pair of items $A$ and $B$, their temporal similarity is defined as:
$$J(Y_A, Y_B) = \frac{|Y_A \cap Y_B|}{|Y_A \cup Y_B|}$$

\begin{figure}[t] 
    \centering
    \includegraphics[width=\columnwidth]{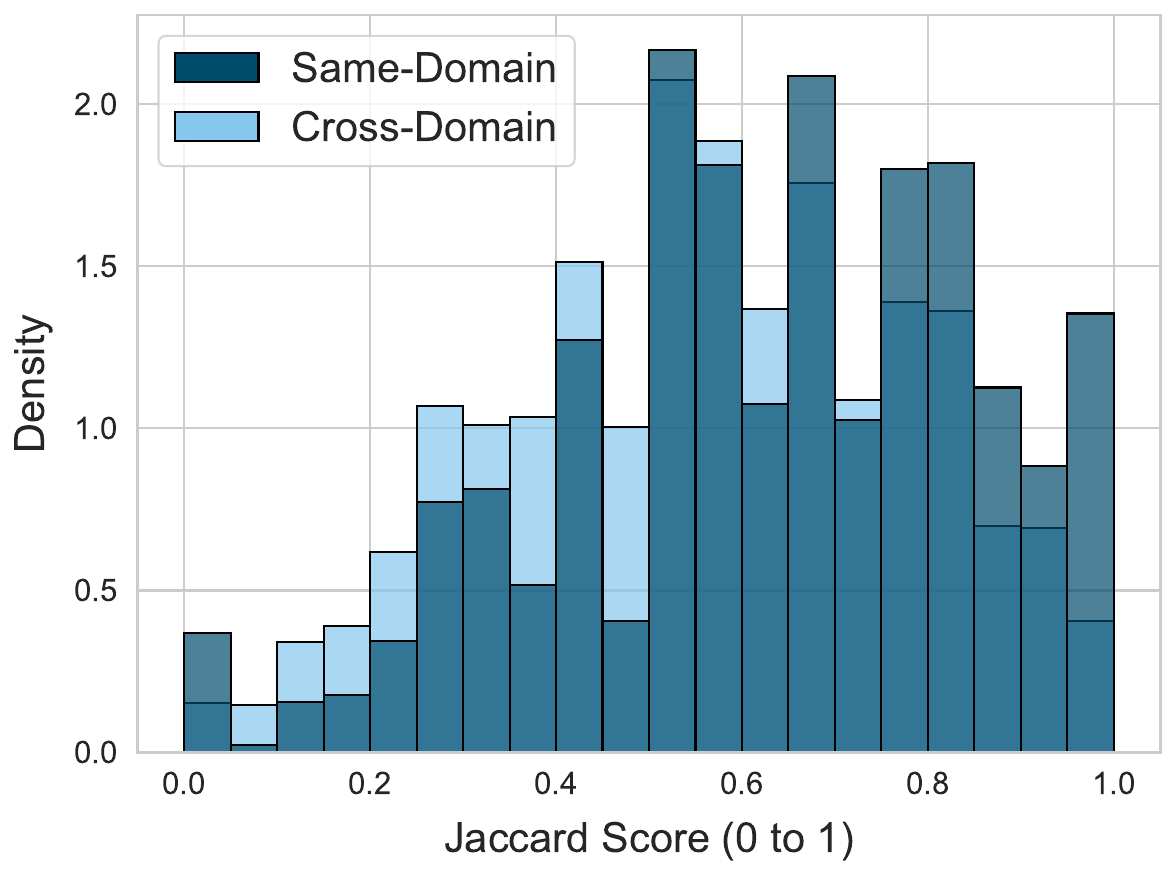} 
    \caption{Density distribution of the temporal Jaccard Index for same-domain and cross-domain item pairs.}
    \label{fig:jaccard_hist}
\end{figure}

A high index indicates contemporaneity. As illustrated in Figure~\ref{fig:jaccard_hist}, both setups share a prominent density concentration above 0.5. However, cross-domain pairs exhibit a visibly wider left-skewed tail, with an increased density between 0.2 and 0.5 compared to the stricter same-domain boundaries. This shared active window mathematically confirms that despite the expected structural divergence across distinct categories, the temporal Jaccard index captures robust contemporaneous alignment, serving as a viable feature for cross-domain knowledge transfer.

\subsubsection{Popularity Metrics}
\begin{figure}[t] 
    \centering
    \includegraphics[width=1\columnwidth]{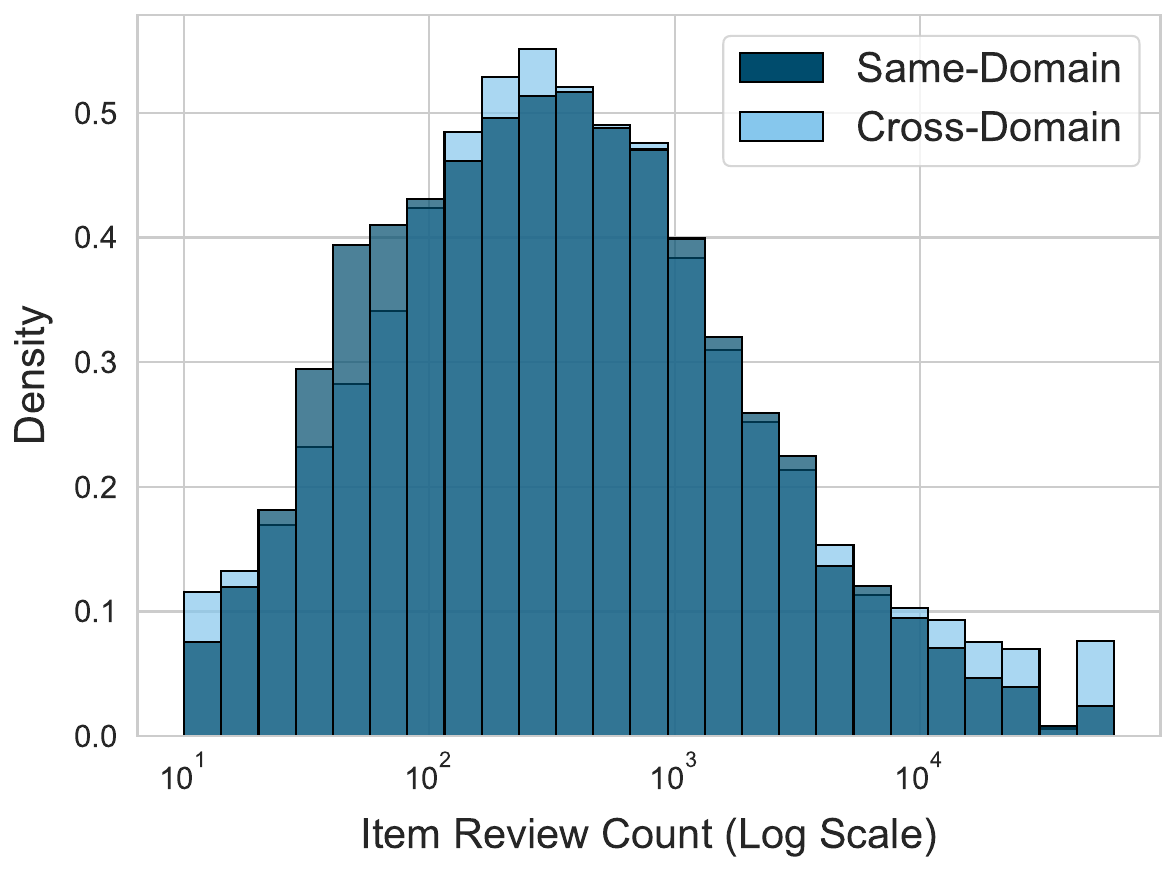} 
    \caption{Density distribution of raw interactions per item.}
    \label{fig:popularity_hist}
\end{figure}

The popularity of an item implies relevant information in similarity learning. To build upon this, we compute the raw review count for each individual item in a given pair ($cnt_A$ and $cnt_B$), and subsequently calculate the absolute difference between them:
\begin{equation}
    \Delta_{popularity} = |cnt_A - cnt_B|
\end{equation}
This pairwise difference acts as a proxy for mainstream appeal alignment. It allows the model to capture whether two items share a similar level of popularity or if there is a severe mismatch in their overall exposure. The individual distribution of these item counts can be seen in Figure \ref{fig:popularity_hist}.

\subsubsection{Ratings Metrics}
We leverage the explicit feedback provided by users through their rating scores (from 1 to 5 stars). For every item pair ($A$, $B$), we calculate the absolute difference between their average user ratings ($\bar{R}_A$ and $\bar{R}_B$):
\begin{equation}
    \Delta_{rating} = |\bar{R}_A - \bar{R}_B|
\end{equation}

\begin{figure}[t] 
    \centering
    \includegraphics[width=1\columnwidth]{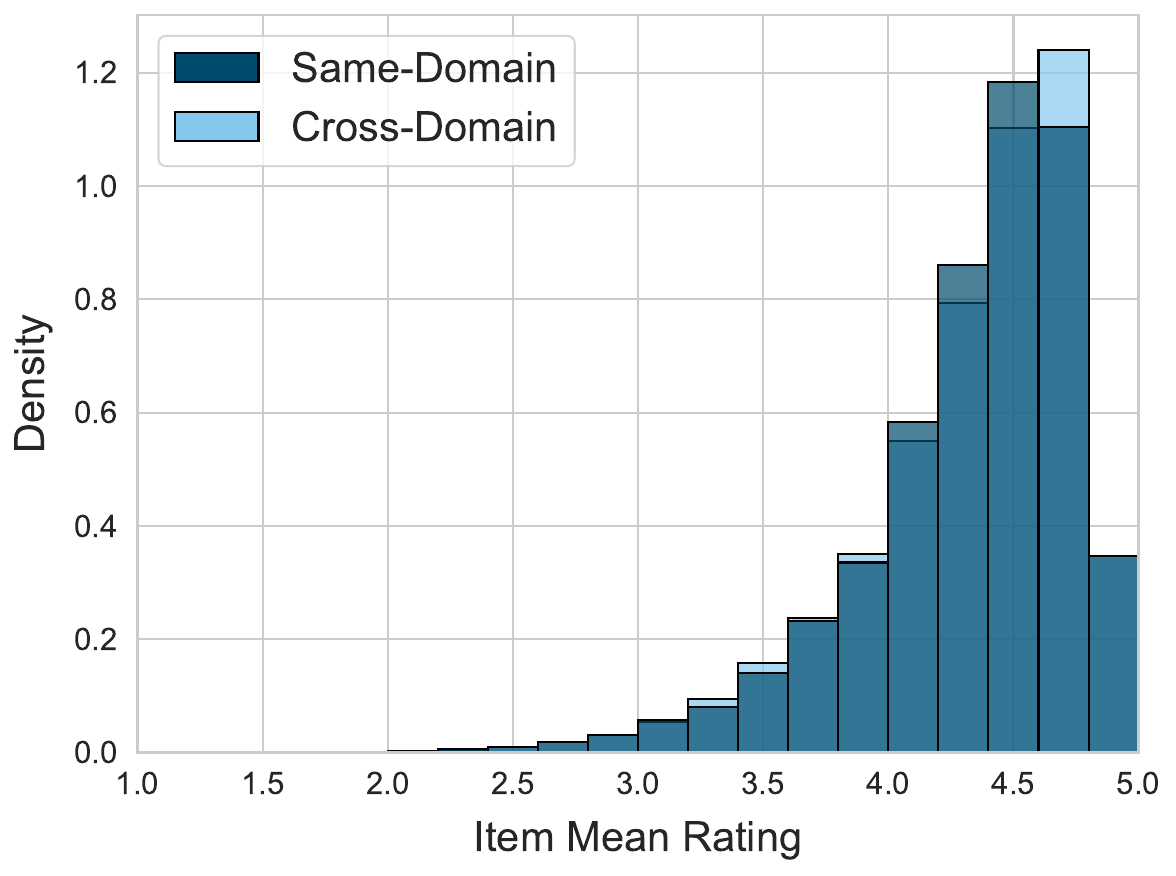} 
    \caption{Distribution of average rating per item.}
    \label{fig:ratings_hist}
\end{figure}

This feature functions as a proxy for quality alignment. It enables the model to identify nuances where two items might be semantically relevant to each other, but possess drastically different levels of user satisfaction, a discrepancy that directly impacts the likelihood of co-occurrence in a user's consumption history. This reliance on explicit rating data as a primary signal for item similarity is well-supported in the literature \cite{koren2009matrix}. The distribution of these individual average scores is presented in Figure~\ref{fig:ratings_hist}.

\subsubsection{Feature Integration}
To evaluate the predictive power of these dimensions, our machine learning framework is designed to ingest these feature groups both in isolation and in a fully integrated manner. Specifically, the high-dimensional SBERT text embeddings are evaluated individually to establish a purely semantic baseline, while the structural metadata (\textit{Jaccard}, reviews count, and rating mean) are modeled separately to assess behavioral trends. Finally, all groups are concatenated into a unified hybrid vector of 775 dimensions:
$$\mathbf{X}_{final} = [\mathbf{E}_A \parallel \mathbf{E}_B \parallel J_{tempo} \parallel cnt_A \parallel cnt_B \parallel \Delta_{pop} \parallel \bar{R}_A \parallel \bar{R}_B \parallel \Delta_{rat}]$$

This modular design allows us to execute ablation studies by systematic removal of feature blocks, isolating the exact performance shifts across our experimental training scenarios.

\section{Experimental Results}
\label{sec:results}

\textbf{Experimental setup.} To evaluate our framework under authentic real-world conditions, our experimental pipeline enforces strict data split protocols, stratification parameters, and baseline contrasts. Prior to any domain filtering or sub-sampling, all shared data pools were divided into an 80/20 ratio for training and testing, ensuring complete isolation to prevent data leakage. To eliminate stochastic variance and guarantee experimental reproducibility, a pseudo-random seed (\texttt{random\_state=42}) was consistently applied across all initializations and iterations. 

While all testing phases were executed strictly on raw, unstratified partitions to preserve authentic long-tail behavior, the training data underwent uniform Random Under-Sampling (RUS) with 100 bins and a maximum of 50,000 pairs per bin to prevent dominance biases, following \cite{pontello2018direction}. Baselines are established via a naive mathematical average predictor (\texttt{DummyRegressor}), which uniformly serves as a comparative bottom-bound. Causal feature importance and domain-shifting dynamics are mapped leveraging game-theoretic frameworks via SHAP \cite{lundberg2017shap} and block-level Feature Ablation Studies \cite{meyes2019ablation}.

\subsection{Cross-Domain Predictive Performance and Feature Ablation}

Our framework is initially evaluated across four distinct operational scenarios structured to measure model resilience against severe domain divergence. These configurations leverage three primary pair categories: \textit{SAME} (intra-domain sets), \textit{CROSS} (heterogeneous book-movie intersections), and \textit{GLOBAL} (the exhaustive union of both environments). 

Specifically, the evaluated environments comprise: 
\textbf{Scenario 1 (CROSS $\rightarrow$ CROSS)}, a specialized model trained with 253,218 pairs and tested on 270,796 raw pairs; 
\textbf{Scenario 2 (GLOBAL $\rightarrow$ GLOBAL)}, a unified model operating under a mixed environment trained with 1,093,562 pairs and tested on 775,796 raw pairs; 
\textbf{Scenario 3 (GLOBAL $\rightarrow$ CROSS)}, which isolates the cross-domain generalization capacity of the global model by testing strictly on the 270,796 CROSS pairs; and 
\textbf{Scenario 4 (SAME $\rightarrow$ CROSS)}, a pure zero-shot domain transfer setup where a model trained exclusively on 949,091 intra-domain pairs is challenged to decipher 1,353,976 cross-domain similarities. 

To prioritize synthesis and focus visual impact on our framework's capabilities, the naive baseline metrics—which yielded extreme negative performance bounds throughout all scenarios—are omitted from the main matrices. Table~\ref{tab:main_results_xgboost} details the empirical results obtained by the proposed hybrid architecture.

\begin{table}[t] 
\centering
\caption{\texttt{Hybrid XGBoost} model: Predictive performance across four operational scenarios.}
\label{tab:main_results_xgboost}
\begin{tabular}{lccc}
\toprule
\textbf{Scenario (Train $\rightarrow$ Test)} & \textbf{$R^2$} & \textbf{RMSE} & \textbf{MAE} \\
\midrule
1. CROSS $\rightarrow$ CROSS & 0.96 & 0.0032 & 0.0014 \\
2. GLOBAL $\rightarrow$ GLOBAL & 0.85 & 0.029 & 0.016 \\
3. GLOBAL $\rightarrow$ CROSS & 0.32 & 0.014 & 0.009 \\
4. SAME $\rightarrow$ CROSS & -4.00 & 0.038 & 0.024 \\
\bottomrule
\end{tabular}
\end{table}

Based on these findings, a stark contrast in predictive performance is observed across the evaluated environments. While the error metrics (RMSE and MSE) present relatively localized variations, the $R^2$ scores reveal critical insights regarding domain shifts. Notably, model performance systematically degrades as the disconnect between the training and testing domains increases. 

In Scenario 3 (\textit{GLOBAL $\rightarrow$ CROSS}), although the model faces a more complex generalization challenge, it still yields a positive $R^2$ score (0.32203). This indicates that embedding the target domain pairs into a heterogeneous global training pool provides sufficient context for the model to capture underlying cross-domain patterns. Conversely, Scenario 4 (\textit{SAME $\rightarrow$ CROSS}) represents a transfer setup that results in a negative $R^2$ score (-4.00211). This demonstrates that the decision rules learned exclusively from intra-domain pairs (\textit{SAME}) fail entirely to generalize to cross-domain interactions (\textit{CROSS}).

\begin{figure}[t]
\centering
\includegraphics[width=1\columnwidth]{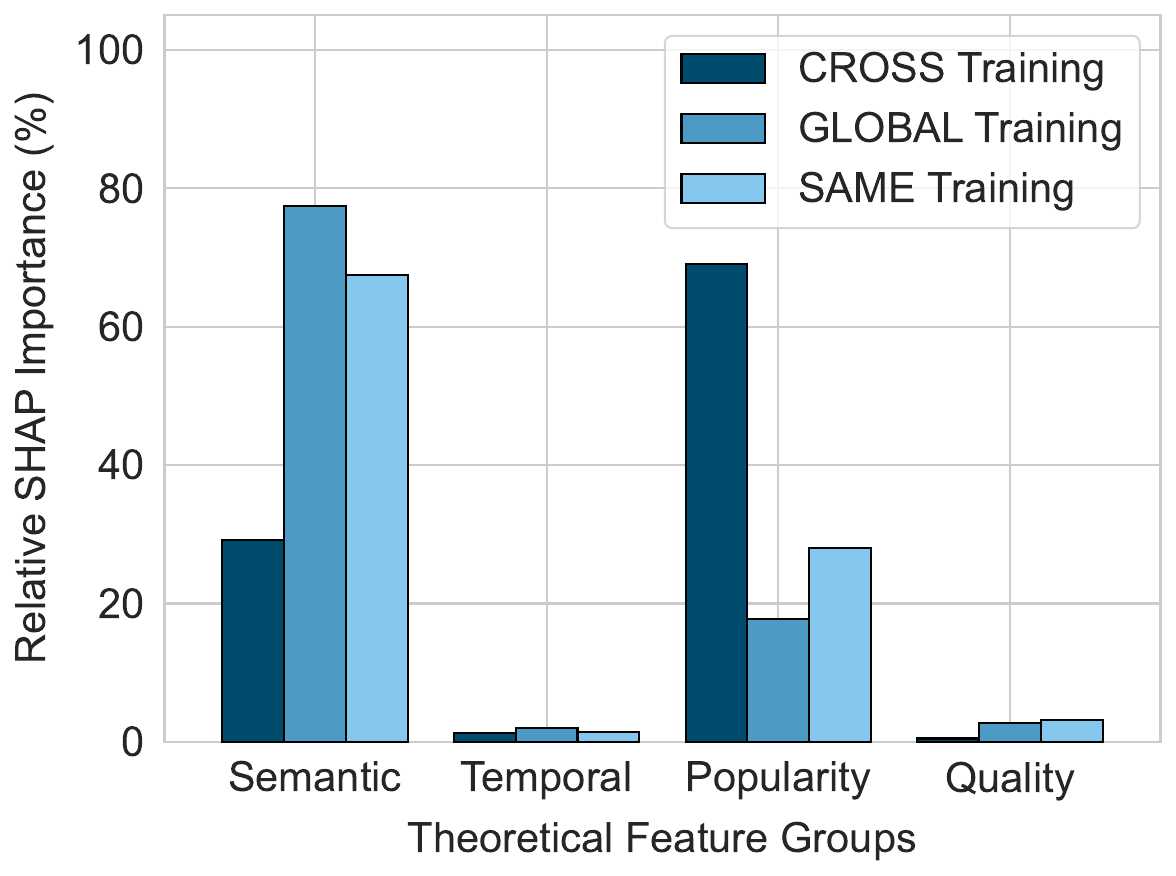}
\caption{Relative SHAP importance shift across training scenarios.}
\label{fig:shap_cross}
\end{figure}

This severe performance asymmetric gap motivated a deeper investigation into the underlying causes of model behavior. Specifically, we leverage SHAP \cite{lundberg2017shap} to analyze whether the models shift their reliance on different feature groups across training setups, which would fundamentally explain this generalization collapse. As illustrated in Figure~\ref{fig:shap_cross}, our hypothesis was empirically validated: the relative importance of the theoretical feature groups undergoes a massive structural shift depending on the training domain. 

In the \textit{SAME Training} setup, the model heavily relies on \textit{Semantic} text features (~68\%) and moderately on \textit{Popularity} (28\%). However, when trained purely on cross-domain data (\textit{CROSS Training}), the model completely flips its strategy, deriving 69\% of its predictive power from \textit{Popularity} behavioral clues and dropping its text \textit{Semantic} reliance to just ~29\%. This mathematically proves why the model trained on \textit{SAME} fails in the zero-shot task: it enters the cross-domain test looking for heavy semantic overlaps that simply do not translate across different categories, whereas the successful cross-model learned that behavioral popularity is the true anchor for cross-domain alignment. 

\paragraph{Feature Ablation Analysis}To provide causal validation for the behavioral shifts identified via SHAP, we conducted a rigorous Feature Ablation Study across both the GLOBAL$\rightarrow$ GLOBAL and  transfer (\textit{GLOBAL$\rightarrow$ CROSS}) training configurations. By systematically isolating and removing each theoretical feature group, we measure their explicit impact on model resilience and generalization.

The ablation results, consolidated in Table~\ref{tab:ablation_global_r2}, empirically substantiate our architectural hypotheses. In the heterogeneous environment (GLOBAL testing), the model heavily relies on structural cues. The removal of \textit{Popularity (Reviews)} triggers a performance collapse, dropping the general $R^2$ from 0.84629 to 0.65455, and causing the transfer performance (\textit{GLOBAL$\rightarrow$ CROSS}) to swing from a positive 0.32203 to a negative -0.98899. A similar critical vulnerability is observed when removing \textit{Semantic (SBERT)} features, where global comes to a $R^2$ of ~0.77 and cross-domain performance plummets to -1.77. Conversely, the ablation of localized metadata like \textit{Temporal (Jaccard)} or \textit{Quality (Ratings)} causes negligible drops, proving that semantic embeddings and interaction volumes form the predictive backbone of cross-domain recommendations.

\begin{table}[t]
\centering
\caption{\texttt{GLOBAL} model: Feature ablation study.}
\label{tab:ablation_global_r2}
\begin{tabular}{lcc}
\toprule
\textbf{Ablated Setup} & \textbf{General Test ($R^2$)} & \textbf{CROSS Test ($R^2$)} \\
\midrule
1. Full Model            & 0.85                          & 0.32                     \\
2. Without Jaccard       & 0.84                          & 0.33                     \\
3. W/out Popularity    & 0.65                          & -0.99                    \\
4. W/out Ratings       & 0.84                          & 0.30                     \\
5. W/out Embeddings    & 0.77                          & -1.78                    \\
\bottomrule
\end{tabular}
\end{table}

In conclusion, these experimental findings demonstrate that the primary challenge lies in executing a zero-shot domain transfer from a homogeneous environment (\textit{SAME}), governed by highly localized feature rules, to a heterogeneous environment (\textit{CROSS}) with distinct underlying dynamics. However, our results reveal that introducing even a modest portion of cross-domain examples into the training phase significantly enhances model adaptability. Despite being disproportionately smaller than the volume of intra-domain data, including these cross-domain pairs effectively bridges the domain sparsity gap. This highlights that incorporating strategic cross-domain context during training is an essential protocol for enabling recommendation frameworks to successfully generalize across diverse operational categories.

\subsection{Intra-Domain Evaluation and Classical Transfer Learning}
\label{sub:intra_transfer}

Given the previous cross-domain limitations, we evaluated the performance of our framework under classical transfer learning regimes, contrasting directional knowledge transfer against the idealized intra-domain upper bounds. To evaluate model scalability beneath traditional domain boundaries, we isolated the \textit{SAME} pool into source-specific intra-domain subsets: Books (762,865 training pairs; 290,175 test pairs) and Movies (331,288 training pairs; 214,826 test pairs). Table~\ref{tab:intra_transfer_results} details the empirical results.

\begin{table}[t]
\centering
\caption{\texttt{Hybrid XGBoost} model: Intra-domain performance and directional classical transfer results.}
\label{tab:intra_transfer_results}
\begin{tabular}{lcc}
\toprule
\textbf{Evaluation Scenario} & \textbf{$R^2$} & \textbf{RMSE} \\
\midrule
Scenario A: Book $\rightarrow$ Book & 0.85 & 0.04 \\
Scenario B: Movie $\rightarrow$ Movie & 0.82 & 0.02 \\
Scenario C: Book $\rightarrow$ Movie & 0.38 & 0.04 \\
Scenario D: Movie $\rightarrow$ Book & 0.76 & 0.05 \\
\bottomrule
\end{tabular}
\end{table}

The empirical results demonstrate strong generalization capabilities, yielding a positive $R^2$ of 0.38 for the Book $\rightarrow$ Movie pipeline and an exceptional $R^2$ of 0.76 for the Movie $\rightarrow$ Book transition. This high performance explicitly confirms that substantial underlying similarities exist across these distinct structural domains. 

\begin{figure}[t]
\centering
\includegraphics[width=1\columnwidth]{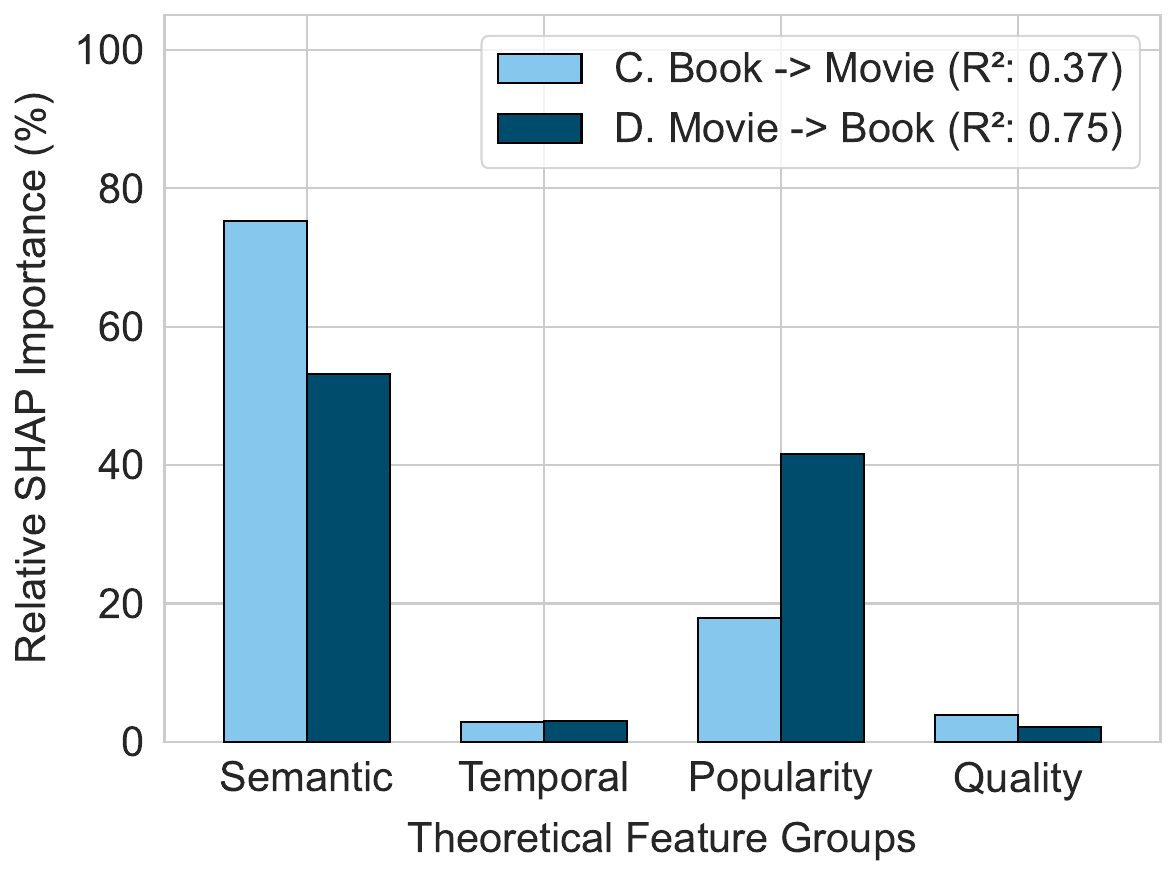}
\caption{Relative SHAP importance shift in domain transfer.}
\label{fig:shap_transfer}
\end{figure}

To map the visual and linguistic alignment driving this behavior, we executed a secondary SHAP feature attribution analysis (Figure~\ref{fig:shap_transfer}). Notably, the transfer models present significantly lower feature discrepancies than those observed in the prior experiments. The asymmetric performance gap between the two transfer tasks can be mathematically explained by the structural variance in semantic reliance. 

The model trained on Books (Book $\rightarrow$ Movie) over-indexes on text \textit{Semantic} cues (75.31\%) while marginalizing behavioral \textit{Popularity} signals (17.94\%). Conversely, the highly successful Movie-to-Book model develops a more balanced strategy, reducing its semantic reliance to 53.18\% and elevating \textit{Popularity} attribution to 41.65\%. This 22\% structural shift proves that knowledge extracted from cinematic consumption patterns builds a more versatile representation, allowing the framework to navigate structural and behavioral dynamics more effectively than a heavily text-centric literary model.

\section{Conclusion and Future Work}
\label{sec:conclusion}

In this work, we presented a scalable supervised framework for cross-domain item similarity learning that successfully operates without relying on proprietary or inaccessible industrial user logs. Our empirical findings demonstrate that while pure zero-shot transfer from homogeneous environments fails due to localized feature biases, introducing even a modest injection of heterogeneous cross-domain pairs dramatically bridges the domain sparsity gap. Furthermore, our game-theoretic feature attribution analysis yielded a critical architectural insight: while intra-domain recommendations heavily index on text-centric semantic embeddings, behavioral popularity metrics serve as the primary foundational anchor for robust, generalized cross-category alignment.

Despite these contributions, our methodology faces constraints regarding data sparsity. The performance drop observed in highly sparse scenarios highlights that when user intersection boundaries are extremely thin, relying strictly on core-filtered co-occurrence frequencies limits the predictive model's ability to capture subtle semantic transitions across vastly different product categories. Consequently, managing this structural sparsity without excessively pruning the interaction graph remains a minor limitation within our current pipeline configuration.

For future research, we intend to address these limitations by exploring alternative metadata from the same data source to serve as additional comparative baselines. We aim to expand our feature engineering pipeline by incorporating richer facets already present in the dataset, such as product categories hierarchy and brand information. Additionally, we plan to validate the generalizability of our architecture by incorporating a music domain and exploring alternative ground-truth metrics, such as the Pearson Correlation Coefficient, to account for explicit user satisfaction curves rather than interaction metadata alone.


\begin{acks}
The authors would like to thank the Coordenação de Aperfeiçoamento de Pessoal de Nível Superior - Brasil (CAPES) - Finance Code 001, the Conselho Nacional de Desenvolvimento Científico e Tecnológico (CNPq), and the Fundação de Amparo à Pesquisa do Estado de Minas Gerais (FAPEMIG) for their financial support.
\end{acks}
\newpage
\bibliographystyle{ACM-Reference-Format}
\bibliography{references_crossdomain}

@STRING{computer = {IEEE Computer}}

@STRING{kdd = {Proceedings of the ACM SIGKDD International Conference on Knowledge Discovery and Data Mining}}

@STRING{sigmod = {Proceedings of the ACM SIGMOD International Conference on Management of Data Conference}}

@STRING{www = {Proceedings of the International World Wide Web Conferences}}

@INPROCEEDINGS{SarwarKKR01,
  author = {Badrul Sarwar and George Karypis and Joseph Konstan and John Riedl},
  title = {{Item-based collaborative filtering recommendation algorithms}},
  booktitle = www,
  year = {2001},
  pages = {285--295},
  address = {Hong Kong, China}
}

@INPROCEEDINGS{reimers2019sentence,
  author = {Nils Reimers and Iryna Gurevych},
  title = {{Sentence-BERT: Sentence Embeddings using Siamese BERT-Networks}},
  booktitle = {{Proceedings of the 2019 Conference on Empirical Methods in Natural Language Processing and the 9th International Joint Conference on Natural Language Processing (EMNLP-IJCNLP)}},
  year = {2019},
  pages = {3982--3992},
  address = {Hong Kong, China}
}

@mastersthesis{pontello2018direction,
  title={Direction-Based Navigation and Similarity Learning across Multiple Media Domains},
  author={Pontello, Luciana Fujii},
  year={2018},
  school={Universidade Federal de Minas Gerais (UFMG)},
  type={Disserta{\c{c}}\~ao de Mestrado},
  address={Belo Horizonte, MG, Brasil}
}

@MISC{hou2024bridging,
  author = {Yupeng Hou and Jiacheng Li and Zhankui He and An Yan and Xiusi Chen and Julian McAuley},
  title = {{Bridging Language and Items for Retrieval and Recommendation}},
  howpublished = {arXiv preprint arXiv:2403.03952},
  year = {2024}
}

@INPROCEEDINGS{raasveldt2019duckdb,
  author = {Mark Raasveldt and Hannes M{\"u}hleisen},
  title = {{DuckDB: an embeddable analytical database}},
  booktitle = sigmod,
  year = {2019},
  pages = {1981--1984},
  address = {Amsterdam, Netherlands}
}

@ARTICLE{linden2003amazon,
  author = {Greg Linden and Brent Smith and Jeremy York},
  title = {{Amazon.com recommendations: Item-to-item collaborative filtering}},
  journal = {IEEE Internet Computing},
  year = {2003},
  volume = {7},
  number = {1},
  pages = {76--80}
}

@INPROCEEDINGS{v2025contrastive,
  author = {Hoang V. Dong and Yuan Fang and Hady W. Lauw},
  title = {{A contrastive framework with user, item and review alignment for recommendation}},
  booktitle = {{Proceedings of the 18th ACM International Conference on Web Search and Data Mining}},
  year = {2025},
  pages = {117--126},
  address = {Hannover, Germany}
}

@article{zhang2025crossdomain,
  title   = {A Comprehensive Survey on Cross-Domain Recommendation: Taxonomy, Progress, and Prospects},
  author  = {Zhang, Hao and Cheng, Mingyue and Liu, Qi and Jiang, Junzhe and Wang, Xianquan and Zhang, Rujiao and Lei, Chenyi and Chen, Enhong},
  journal = {arXiv preprint arXiv:2503.14110},
  year    = {2025}
}

@inproceedings{zhao2023crossdomain,
  title     = {Cross-domain recommendation via user interest alignment},
  author    = {Zhao, Chuang and Zhao, Hongke and He, Ming and Zhang, Jian and Fan, Jianping},
  booktitle = {Proceedings of the ACM Web Conference 2023 (WWW '23)},
  pages     = {887--896},
  year      = {2023}
}

@inproceedings{cao2023universal,
  title     = {Towards universal cross-domain recommendation},
  author    = {Cao, Jiangxia and Li, Shen and Yu, Bowen and Guo, Xiaobo and Liu, Tingwen and Wang, Bin},
  booktitle = {Proceedings of the Sixteenth ACM International Conference on Web Search and Data Mining (WSDM '23)},
  pages     = {78--86},
  year      = {2023}
}

@inproceedings{zhao2024discerning,
  title     = {Discerning Canonical User Representation for Cross-Domain Recommendation},
  author    = {Zhao, Siqian and Sahebi, Sherry},
  booktitle = {Proceedings of the 18th ACM Conference on Recommender Systems (RecSys '24)},
  pages     = {318--328},
  year      = {2024}
}

@inproceedings{zheng2017joint,
  title     = {Joint deep modeling of users and items using reviews for recommendation},
  author    = {Zheng, Lei and Noroozi, Vahid and Yu, Philip S.},
  booktitle = {Proceedings of the Tenth ACM International Conference on Web Search and Data Mining (WSDM '17)},
  pages     = {425--434},
  year      = {2017}
}

@article{dacrema2021troubling,
  title   = {A Troubling Analysis of Reproducibility and Progress in Recommender Systems Research},
  author  = {Ferrari Dacrema, Maurizio and Boglio, Simone and Cremonesi, Paolo and Jannach, Dietmar},
  journal = {ACM Transactions on Information Systems},
  volume  = {39},
  number  = {2},
  pages   = {20:1--20:49},
  year    = {2021},
  doi     = {10.1145/3434185}
}

@article{zou2025survey,
  title   = {A Survey of Real-World Recommender Systems: Challenges, Constraints, and Industrial Perspectives},
  author  = {Zou, Kuan and Sun, Aixin},
  journal = {arXiv preprint arXiv:2509.06002},
  year    = {2025}
}

@inproceedings{alonsojimenez2022music,
  title     = {Music Representation Learning Based on Editorial Metadata from Discogs},
  author    = {Alonso-Jim{\'e}nez, Pablo and Serra, Xavier and Bogdanov, Dmitry},
  booktitle = {Proceedings of the 23rd International Society for Music Information Retrieval Conference (ISMIR 2022)},
  pages     = {825--833},
  year      = {2022},
  doi       = {10.5281/zenodo.7316790}
}

@article{liu2025amazon,
  title   = {Building a Recommendation System Using Amazon Product Co-Purchasing Network},
  author  = {Liu, Mark and Zhao, Catherine and Zhou, Nathan},
  journal = {arXiv preprint arXiv:2506.02482},
  year    = {2025}
}

@article{linden2003item,
  title   = {Amazon.com recommendations: Item-to-item collaborative filtering},
  author  = {Linden, Greg and Smith, Brent and York, Jeremy},
  journal = {IEEE Internet Computing},
  volume  = {7},
  number  = {1},
  pages   = {76--80},
  year    = {2003}
}

@article{koren2009matrix,
  title   = {Matrix factorization techniques for recommender systems},
  author  = {Koren, Yehuda and Bell, Robert and Volinsky, Chris},
  journal = {Computer},
  volume  = {42},
  number  = {8},
  pages   = {30--37},
  year    = {2009}
}

@inproceedings{ni2019justifying,
  title     = {Justifying Recommendations using Distantly-Labeled Reviews and Fine-Grained Aspects},
  author    = {Ni, Jianmo and Li, Jiacheng and McAuley, Julian},
  booktitle = {Proceedings of the 2019 Conference on Empirical Methods in Natural Language Processing and the 9th International Joint Conference on Natural Language Processing (EMNLP-IJCNLP)},
  pages     = {188--197},
  year      = {2019}
}

@inproceedings{23webmediaMultidomain,
 author = {Alexandre Bender and Emillyn Souza and Ihan Bender and Ulisses Corrêa and Ricardo Araujo},
 title = { Improving Multi-Domain Learning by Balancing Batches With Domain Information},
 booktitle = {Proceedings of the 29th Brazilian Symposium on Multimedia and the Web},
 location = {Ribeirão Preto/SP},
 year = {2023},
 issn = {0000-0000},
 pages = {96–103},
 publisher = {SBC},
 address = {Porto Alegre, RS, Brasil},
 url = {https://sol.sbc.org.br/index.php/webmedia/article/view/25870}
}

@inproceedings{25webmediacalibration,
 author = {Atauchi, Paul and Zani, André Levi and Rocha, Leonardo and Manzato, Marcelo G.},
 title = { Aprimorando a Personalização de Sistemas de Recomendação por Calibração Multicategoria},
 booktitle = {Proceedings of the 31st Brazilian Symposium on Multimedia and the Web},
 location = {Rio de Janeiro/RJ},
 year = {2025},
 issn = {0000-0000},
 pages = {},
 publisher = {SBC},
 address = {Porto Alegre, RS, Brasil},
 url = {https://sol.sbc.org.br/index.php/webmedia/article/view/37998}
}

@inproceedings{2009korentemporal,
    author = {Koren, Yehuda},
    title = {Collaborative filtering with temporal dynamics},
    booktitle = {KDD '09: Proceedings of the 15th ACM SIGKDD international conference on Knowledge discovery and data mining},
    year = {2009}
}

@article{meyes2019ablation,
  title={Ablation studies in artificial neural networks},
  author={Meyes, Richard and Chen, Lu and de Puiseau, Constantin and Meisen, Tobias},
  journal={arXiv preprint arXiv:1901.08644},
  year={2019}
}

@incollection{lundberg2017shap,
  title={A unified approach to interpreting model predictions},
  author={Lundberg, Scott M and Lee, Su-In},
  booktitle={Advances in Neural Information Processing Systems},
  volume={30},
  year={2017},
  publisher={Curran Associates, Inc.}
}

@book{manning2008tfidf,
  title={Introduction to information retrieval},
  author={Manning, Christopher D and Raghavan, Prabhakar and Sch{\"u}tze, Hinrich},
  year={2008},
  publisher={Cambridge university press},
  address={Cambridge, UK}
}


\end{document}